\documentclass[prb,onecolumn,superscriptaddress,preprintnumbers,amssymb]{revtex4-2}
\usepackage{graphicx}
\usepackage{color}
\usepackage{dcolumn}
\usepackage{bm}
\usepackage{graphicx}
\usepackage{amsmath}
\usepackage{array}
\usepackage{ulem}
\usepackage{hyperref}
\usepackage{comment}
\usepackage{slashed}
\newcommand{\dd}{\mathrm{d}}

\newcommand{\beq}{\begin{equation}}
\newcommand{\eeq}{\end{equation}}
\newcommand{\beqn}{\begin{eqnarray}}
\newcommand{\eeqn}{\end{eqnarray}}

\newcommand{\ra}{\rightarrow}

\newcommand{\vect}[1]{{\bm{#1}}}

\newcommand{\ii}{\mathrm{i}}

\newcommand{\vE}{\vect{E}}

\newcommand{\vA}{\vect{A}}

\newcommand{\U}{\mathrm{U}}

\newcommand{\cx}[1]{{\color{black} #1}}
\newcommand{\cxa}[1]{{\color{black} #1}}

\begin{document}

\title{Particle-Vortex Duality of Hydrodynamics}

\author{Cenke Xu}

\author{Matthew P. A. Fisher}

\affiliation{Department of Physics, University of California, Santa Barbara, CA 93106, USA}

\date{\today}

\begin{abstract}

Equipped with the recently recognized symmetry structure of mixed states of matter and ``strong-weak spontaneous symmetry breaking" (SW-SSB), we develop a quantum particle-vortex duality of emergent model-F and model-A hydrodynamics of $2d$ boson/rotor systems. This duality relation demonstrates that classical hydrodynamics of $2d$ bosons can be described in terms of the charge symmetry $\U(1)_c$, but also equivalently in terms of the dual (emergent) 1-form symmetry $\U(1)^{(1)}_e$, as well as $\U(1)_v$ associated with the conservation of vortices. This duality provides a bridge between the hydrodynamics of matter and magnetohydrodynamics. 

\end{abstract}

\maketitle

\section{Introduction}

Hydrodynamics is among the oldest theoretical frameworks developed to describe nature. It is also the ``common language" across different disciplines of science, describing the transport of strongly correlated microscopic particles, such as electrons in condensed matter and quark-gluon matter in QCD, as well as the collective behavior of active matter, such as bird flocking, and even dynamics on galactic scales in astrophysics. All these hydrodynamics share one common feature: information of each individual object, such as charge quantization and spatial position, is unimportant or irrelevant for the collective behaviors at long distance, therefore physics in the infrared can be well captured by equations with continuous variables, such as particle number and momentum density. 

Duality is another celebrated organizing principle in physics. It refers to the situation in which two, or sometimes several, theories that appear entirely different nevertheless describe the same underlying physics. Degrees of freedom that are elementary particles in one formulation may emerge as topological defects in another, while both descriptions encode the same physical content. Prominent examples include particle-vortex duality in condensed-matter physics~\cite{peskindual,halperindual,leefisherdual} as well as the duality relating different superstring theories. More recently, a duality web has been proposed that connects a class of exotic quantum critical points in $(2+1)d$~\cite{SO5,dualreview}.

\cxa{Duality has traditionally been explored primarily in equilibrium systems, and only recently has it been extended to nonequilibrium settings~\cite{piotr}}. In the current paper, \cxa{starting with full quantum systems on the lattice}, we explore duality between different non-equilibrium quantum and emergent hydrodynamic systems, 
building on the recent discussion of the so-called ``strong-weak spontaneous symmetry breaking" (SW-SSB)~\cite{Lee_2023,Ma_2023,wangSWSSB,Chen_2025,weinstein2025efficient,wangreview,SWSSBex}, and in particular, its implications for the emergence of classical hydrodynamics from underlying quantum dynamics~\cite{Ogunnaike2023hydro,huang2025hydrodynamics,DelacretazBSS2025HydroEFT,wangSWSSB,hauser2026}. In particular, as was clarified in Ref.~\cite{hauser2026}, after the SW-SSB of the charge-U(1) symmetry in Lindbladian dynamics, the discreteness of charge can no longer be distinguished, and the underlying charge density becomes a continuous variable, a necessary condition for the emergence of hydrodynamics.  
By employing quantum Lindbladian dynamics, our results show that different hydrodynamics discussed by Hohenberg-Halperin~\cite{HH} may in fact be dual to one another. In particular, we find that \uline{``model-F" dynamics of one symmetry arising from strong-weak spontaneous symmetry breaking (SW-SSB) could be dual to the ``model-A" dynamics of a dual symmetry arising from the weak-trivial spontaneous symmetry breaking (WT-SSB)}. 

Let us first describe our central results of Section~\ref{sec:rotor_modelF_modelA_review} and Section~\ref{2dPVduality}, 
which we briefly summarize in Table I. The target system of Section~\ref{sec:rotor_modelF_modelA_review} is a $2d$ quantum boson/rotor system, and we will discuss three symmetries: the charge-U(1) symmetry $\U(1)_c$ associated with the rotor number conservation, the $\U(1)_v$ symmetry associated with vortex number conservation, and the dual electric 1-form symmetry $\U(1)^{(1)}_e$. Note that $\U(1)^{(1)}_e$ can only be emergent due to the existence of vortices (matter fields coupled to the 1-form gauge field). 

First of all, when our Lindblad dynamics has a strong $\U(1)_c$ symmetry, three different phases may arise: 

\begin{itemize}

\item The Mott insulator phase, which is the phase with strongly symmetric $\U(1)_c$. Under the particle-vortex duality, the Mott insulator is also a phase where the vortices condense, therefore the Mott insulator exhibits strong-trivial spontaneous symmetry breaking (ST-SSB) of $\U(1)_v$, while $\U(1)^{(1)}_e$ is strongly broken explicitly and does not emerge in the IR. 

\item If we initialize the system in the Mott insulator phase, under {\it finite-time} decoherence dynamics the system can enter a normal fluid phase, via a strong-to-weak spontaneous symmetry breaking transition of the $\U(1)_c$ symmetry. This is the phase where classical model-F hydrodynamics starts to emerge, and the characteristic signature of this phase is charge diffusion (Sec.~\ref{normal}). The normal fluid phase exhibits SW-SSB of both $\U(1)_c$ and $\U(1)_v$ symmetries (Sec.~\ref{normal} and Sec.~\ref{dualnormal}, respectively). There is also an emergent weak-$\U(1)^{(1)}_e$, which undergoes W$^\ast$T-SSB, where the asterisk refers to the emergence of the symmetry. Thereby we establish a duality between the hydrodynamics consequences of 0-form SW-SSB (model-F) and 1-form WT-SSB (model-A). 

\item Under further decoherence dynamics, the system can enter a superfluid phase (Sec.~\ref{super} and Sec.~\ref{dualsuper}). We note that it must take infinite time to reach the superfluid from the normal fluid phase, as the linear-density matrix correlation becomes long-ranged in the superfluid, which is prohibited from happening at finite-time due to the Lieb-Robinson bound. When there is only non-unitary decoherence dynamics, both the charge density and order parameter (phase field) diffuse, 
and they intertwine into a single Goldstone mode under additional unitary Hamiltonian dynamics. This is the phase with ST-SSB of $\U(1)_c$, and strongly symmetric $\U(1)_v$. There is also an emergent strong-$\U(1)^{(1)}_e$, which undergoes S$^\ast$T-SSB. Therefore the ST-SSB of 0-form symmetry is dual to the ST-SSB of 1-form symmetry, consistent with the previously established duality in equilibrium. 

\end{itemize}

Secondly, we can also explicitly break the strong-$\U(1)_c$ symmetry to weak-$\U(1)_c$ symmetry by including single boson creation/annihilation operators as Lindblad jump operators. In this case two phases can arise: 

\begin{itemize}

\item The normal fluid without diffusion, as the diffusion is suppressed by the weakly-symmetric jump operators. This phase also has an emergent-weak-$\U(1)^{(1)}_e$ symmetry, and it is not spontaneously broken. 

\item The superfluid phase with diffusion of the phase field. This phase exhibits WT-SSB of the $\U(1)_c$ symmetry, and is strongly symmetric under the $\U(1)_v$ symmetry. There is also an emergent strong-$\U(1)^{(1)}_e$, and it undergoes S$^\ast$W-SSB. Therefore there is a duality between 0-form WT-SSB (model-A), and also 1-form SW-SSB (model-F). 

\end{itemize}

\begin{table}[h]
\centering
\scriptsize
\renewcommand{\arraystretch}{2.5}
\begin{tabular}{|l|l|l|l|l|l|}
\hline
Dynamics & Phase & $\U(1)_c$ & $\U(1)_v$ & $\U(1)^{(1)}_e$ & Hydro modes \\
\hline

Model-F, strong $\U(1)_c$
&
Mott insulator
&
Strongly Symmetric
&
ST-SSB
&
Strongly Broken
&
none
\\
\hline

Model-F, strong $\U(1)_c$
&
Normal fluid
&
SW-SSB
&
SW-SSB
&
W$^\ast$T-SSB
&
diffusion of $n$ (dual $B$)
\\
\hline

Model-F, strong $\U(1)_c$
&
Superfluid
&
ST-SSB
&
Strongly Symmetric
&
S$^\ast$T-SSB
&
Goldstone mode
\\
\hline

Model-A, weak $\U(1)_c$
&
Normal fluid
&
Weakly Symmetric
&
SW-SSB
&
Weakly$^\ast$ Symmetric
&
none
\\
\hline

Model-A, weak $\U(1)_c$
&
Superfluid
&
WT-SSB
&
Strongly Symmetric
&
S$^\ast$W-SSB
&
diffusion of $\phi$ (dual $E_T$)
\\
\hline

\end{tabular}
\caption{Symmetry structure and hydrodynamic modes in the particle-vortex dual description of rotor hydrodynamics. Here `` $^\ast$ " denotes an emergent symmetry.}
\label{tab:PVduality_symmetry_hydro}
\end{table}

In the appendix, we also discuss the nonunitary dynamics of a $2d$ compact U(1) gauge theory with an explicit strong or weak 1-form symmetry. We verify the following: 

\begin{itemize}

\item In $2d$, 0-form SW-SSB with model-F dynamics is dual to 1-form WT-SSB with model-A dynamics; see Sec.~\ref{2dcompactgaugeSWSSB}. Conversely, 1-form SW-SSB with model-F dynamics is dual to 0-form WT-SSB with model-A dynamics; see Sec.~\ref{2dcompactgaugeWTSSB}.

\end{itemize}
 
\section{Emergence of model-F and model-A hydrodynamics}
\label{sec:rotor_modelF_modelA_review}

In this section we review how the long-distance classical hydrodynamic equations arise from a microscopic quantum rotor model under decoherence dynamics, see ~\cite{hauser2026}. We use $\phi_i$ for the rotor phase and $n_i$ for the rotor number,
\beqn
[\hat{\phi}_i,\hat{n}_j]=\ii\delta_{ij},
\eeqn
with $i,j$ being site labels for the 2d square lattice.
The charge symmetry is $\U(1)_c: \phi_i\rightarrow \phi_i+\alpha$.
We will take Lindbladian dissipative dynamics to satisfy detailed balance with respect to the classical free energy,
\beqn
F[n,\phi]=F_n[n]+F_\phi[\phi]
={K\over2}\sum_i n_i^2-J\sum_{\langle i,j\rangle}\cos(\phi_i-\phi_j).
\label{eq:secII_rotor_F}
\eeqn
When quantum coherent dynamics is included, we allow a Hamiltonian which have independent coefficients,
\beqn
\hat{H}[n,\phi]={u_n\over2}\sum_i \hat{n}_i^2-u_\phi\sum_{\langle i,j\rangle}\cos(\hat{\phi}_i-\hat{\phi}_j).
\label{eq:secII_rotor_H}
\eeqn
One may set $u_n=K$ and $u_\phi=J$ to recover the simplest choice $H=F$.

We will express the Lindbladian dynamics in terms of a Keldysh path integral, see~\cite{hauser2026}, with forward/backward fields, $\phi_{R/L}$ and $n_{R/L}$.  It is convenient to introduce the Keldysh classical and quantum fields as,
\beqn
\phi={\phi_R+\phi_L\over2}, \qquad \tilde\phi=\phi_R-\phi_L, \qquad n={n_R+n_L\over2},\qquad \tilde n=n_R-n_L.
\eeqn
The kinematic part of the Keldysh action is
\beqn
S_0= \ii \int \dd t \sum_i \ ( n_{R,i}\dot\phi_{R,i} - n_{L,i}\dot{\phi}_{L,i}) = \ii \int \dd t \sum_i \ ( \tilde n_i\dot\phi_i+n_i\dot{\tilde\phi}_i ).
\label{eq:secII_rotor_S0}
\eeqn
Notice that the quantum phase field, $\tilde{\phi}$, is canonically conjugate to the classical number field, $n$, and vice versa.
The strong symmetry shifts the quantum phase field $\tilde\phi$, while the weak symmetry shifts the classical phase field $\phi$:
\beqn
\U(1)_{c,{\rm strong}}:\quad \tilde\phi\rightarrow \tilde\phi+\alpha,
\qquad
\U(1)_{c,{\rm weak}}:\quad \phi\rightarrow \phi+\alpha.
\eeqn
Thus long-range or quasi-long-range order of $e^{\ii\tilde\phi}$ diagnoses SW-SSB of $\U(1)_c$. Once this happens, $\tilde\phi$ may be treated as a spin-wave field. Since $n$ is conjugate to $\tilde\phi$, this is precisely the step where the microscopic quantization of $n$ is lost and $n$ becomes a continuous hydrodynamic variable.

\subsection{Strong-$\U(1)_c$ jumps and the model-F hydrodynamics}

We first consider Lindblad jump operators preserving the strong $\U(1)_c$ symmetry. The first jump operator shifts the rotor phase on a site,
\beqn
\hat L_{\phi,i}^{a}= \sqrt{\gamma_\phi} \exp(-\ii a\hat n_i) \exp\left[-{\beta\over4}\left(F_\phi[\hat\phi_i+a]-F_\phi[\hat\phi_i]\right)\right], \qquad a\in\mathbb R.
\label{eq:secII_Lphi}
\eeqn
and we will consider the $a \rightarrow 0$ limit~\cite{hauser2026}.  The corresponding term in the Lindbladian is,
\begin{equation}
 {\cal L}_\phi[\hat{\rho}] =  \sum_j \int_a \partial_a^2 \delta(a) [ \hat{L}^a_{\phi_j}\hat{\rho}  \hat{L}^{a \dagger}_{\phi_j} - \frac{1}{2} \{ \hat{L}^{a \dagger}_{\phi_j} \hat{L}^a_{\phi_j} , \hat{\rho}\} ].
 \end{equation}

The second jump moves one unit of charge across a nearest-neighbor link,
\beqn
\hat L_{n,ij}^{s}= \sqrt{\gamma_n} \exp\left[\ii s(\hat\phi_i-\hat\phi_j)\right] \exp\left[-{\beta\over4}\left(F_n[\hat n_i+s,\hat n_j-s]-F_n[\hat n_i,\hat n_j]\right)\right], \qquad s=\pm1.
\label{eq:secII_Ln}
\eeqn
The corresponding contribution to the Lindbladian is,
\begin{equation}
 {\cal L}_n [\hat{\rho}] = \sum_{\langle i, j \rangle} \sum_{\pm}  [ \hat{L}^\pm_{n_{ij}}\hat{\rho}  \hat{L}^{\pm \dagger}_{n_{ij}} - \frac{1}{2} \{ \hat{L}^{\pm \dagger}_{n_{ij}} \hat{L}^\pm_{n_{ij}} , \hat{\rho}\} ].
\end{equation}

The full Lindblad equation of motion is taken as,
\begin{equation}
\partial_t \hat{\rho} = - \ii [ \hat{H}, \hat{\rho} ] + {\cal L}_\phi [\hat{\rho}] + {\cal L}_n [\hat{\rho}].
\end{equation}
This dynamics has a strong $\U(1)_c$ symmetry since the total charge $\hat{Q}=\sum_i \hat{n}_i$
commutes with the Hamiltonian as well as both sets of jump operators.

In the Lindblad dynamics these jumps operators generate terms in the Keldysh action, ~\cite{hauser2026},
\beqn
S_\phi=\gamma_\phi\int\dd t\sum_i -\tilde n_i^2 -\ii\beta J\tilde n_i\sum_{j\in i} \left(\sin(\phi_{Rj}-\phi_{Ri}) + \sin(\phi_{Lj}-\phi_{Li})\right) ,
\label{eq:secII_Sphi}
\eeqn
and
\beqn
S_n=\gamma_n\int\dd t\sum_{\langle i,j\rangle}  2\cos(\tilde\phi_i-\tilde\phi_j) -\ii\beta K(n_i-n_j)\sin(\tilde\phi_i-\tilde\phi_j) .
\label{eq:secII_Sn}
\eeqn
Here, we are working to leading order in $\beta J$ and $\beta K$.

In the rest of this section we spell out three phases of the decohered $2d$ quantum rotor.

\subsubsection{Mott insulator phase} \label{Mott}

At time $t=0$ we start in a pure state with $n_i=0$ on each site - this is a Mott insulator.  We then turn on the Lindbladian dissipative dynamics.  At early times we stay in a mixed state Mott-insulator,
where the charge is locally quantized and the rotor phase is strongly disordered. In the Keldysh language, $\tilde\phi$ is not in a Gaussian spin-wave regime, so the Mott insulator is the strongly symmetric phase of $\U(1)_c$. Therefore one should not expand the compact functions (cosine and sine functions) in Eq.~\ref{eq:secII_Sn}.  Equivalently, the microscopic discreteness of $n_i$ has not been washed out. In particular, the Mott phase has no hydrodynamic mode of $\phi$ or $n$.

\subsubsection{Normal fluid phase} \label{normal}

We next consider the normal fluid phase. As was shown in Ref.~\cite{hauser2026}, for a $2d$ rotor model under decoherence, there is a finite critical time $t_c$ beyond which the initial Mott insulator state develops SW-SSB, i.e. the condensation of $\tilde{\phi}$ (more precisely it is a quasi-long range order of $e^{\ii \tilde{\phi}}$). For $t > t_c$ in $2d$, after the SW-SSB transition, we may perform the polynomial-expansion of the quantum phase field,
\beqn
\cos(\tilde\phi_i-\tilde\phi_j)\simeq 1-{1\over2}(\tilde\phi_i-\tilde\phi_j)^2, \qquad \sin(\tilde\phi_i-\tilde\phi_j)\simeq \tilde\phi_i-\tilde\phi_j.
\eeqn
After introducing Hubbard-Stratonovich noise fields to decouple the terms quadratic in $\tilde{n}$ and $\tilde{\phi}$, the action is linearized. Then, after the SW-SSB transition, the Lindbladian dynamics becomes entirely classical, described in terms of equations of motion for the classical fields, $n$ and $\phi$:
\beqn
\frac{\delta S}{ \delta \tilde{n}} = 0 \quad \ra \quad \dot\phi_i &=& {\partial H\over\partial n_i} - \beta\gamma_\phi {\partial F\over\partial\phi_i} + \eta_i,
\cr\cr
\frac{\delta S}{ \delta \tilde{\phi}} = 0 \quad \ra \quad \dot n_i &=& -{\partial H\over\partial\phi_i} + \beta\gamma_n\nabla_i^2 {\partial F\over\partial n_i} + \nabla\cdot\zeta_i.
\label{eq:secII_modelF_langevin}
\eeqn
Here $\nabla_i^2 f_i=\sum_{j\in i}(f_j-f_i)$,
and the noise is $\delta-$correlated white noise,
\beqn
\overline{\eta_i(t)\eta_j(t')}=2\gamma_\phi\delta_{ij}\delta(t-t'), \qquad \overline{\zeta_\mu(x,t)\zeta_\nu(x',t')}=2\gamma_n\delta_{\mu\nu}\delta(x-x')\delta(t-t').
\eeqn

In the normal fluid phase at high temperature we can set $\phi \approx 0$ and the remaining charge-hopping jump operator gives a hydrodynamic equation for the now-continuous density $n$:
\beqn
\dot n = D_n\nabla^2 n + \nabla \cdot\zeta, \qquad D_n=\beta\gamma_nK.
\label{eq:secII_n_diff_noH}
\eeqn 

Later we will see that \uline{there is a duality between 0-form $\U(1)_c$ SW-SSB (model-F) and 1-form $\U(1)^{(1)}_e$ WT-SSB (model-A).}

\subsubsection{Superfluid phase} \label{super}

Finally we consider the superfluid phase, which is the ST-SSB phase of $\U(1)_c$. In this phase, both $\tilde\phi$ and $\phi$ condense and are amenable to spin-wave expansions. The free energy becomes,
\beqn
F\simeq \int\dd^2x \ \left( {K\over2}n^2+{J_R\over2}(\nabla\phi)^2 \right), \label{freeenergyrotorSF}
\eeqn
where $J_R$ is the renormalized phase stiffness. We first assume that there is no Hamiltonian dynamics, $H = 0$, then Eq.~\ref{eq:secII_modelF_langevin} reduces to the purely incoherent equations:
\beqn
\dot n &=& D_n\nabla^2 n + \nabla \cdot\zeta, \qquad D_n = \beta\gamma_nK, \cr\cr \dot\phi &=& D_\phi\nabla^2 \phi + \eta, \qquad D_\phi = \beta\gamma_\phi J_R,
\label{eq:secII_phi_diff_noH}
\eeqn
where we have taken the spatial continuum limit.
Therefore, without Hamiltonian dynamics, the superfluid has two diffusive hydrodynamic fields $n$ and $\phi$.
Later we will see that \uline{under the particle-vortex duality this double-diffusion is dual to the simultaneous diffusion of magnetic ($B$) and electric ($E_T$) fields}.

Now let us include the coherent Hamiltonian $H$ 
which corresponds to a term in the Keldysh action of the form,
\begin{equation}
    S_{H} = i \int dt \left( H(\phi_R,n_R) - H(\phi_L,n_L) \right).
\end{equation}
After SW-SSB we can expand to linear order in small $\tilde{\phi} = \phi_R - \phi_L$.  The final model-F Langevin equations then become,
\beqn
\partial_t\phi &=& u_n n + D_\phi\nabla^2\phi + \eta, \cr\cr \partial_t n &=& u_\phi\nabla^2\phi + D_n\nabla^2 n +
\nabla\cdot\zeta.
\label{eq:secII_modelF_H_super}
\eeqn
Notice that the Hamiltonian couples the two hydrodynamic fields $n$ and $\phi$. This coupling is only meaningful in the superfluid phase, because only here is $\phi$ a long-wavelength spin-wave variable. At small momentum these equations lead to mode,
\beqn
\omega_\pm(k) = \pm ck - \ii {D_\phi+D_n\over2}k^2 +O(k^3).
\eeqn
Thus the superfluid has a weakly damped propagating Goldstone mode.  \uline{As we shall see in the next Section, in the dual vortex description this becomes a propagating ``photon mode" in terms of the dual gauge field.}

\cx{We note that if one starts with an initial crystalline state of particles (in the spatial continuum, say), intuitively melting of the crystal might be viewed as the onset of hydrodynamics. But crystal melting is fundamentally different from the SW-SSB discussed here. Crystal melting can be diagnosed through linear-density-matrix correlations, such as Bragg structure factor. And if one starts from a crystalline state, it can only occur after infinite-time Lindbladian evolution, consistent with the Lieb–Robinson bound. In contrast, SW-SSB can occur inside the crystal phase (due to vacancies and interstitials) at finite times after turning on the dissipation, and is diagnosed by nonlinear-density-matrix correlations, such as a Rényi correlator. }

\subsection{Weak symmetry and model-A dynamics} \label{modelA}

We now explicitly break the strong $\U(1)_c$ down to weak $\U(1)_c$ by adding an onsite charge-changing Lindblad jump operator,
\beqn
\hat L_{w,i}^{s}= \sqrt{\gamma_w} \exp(\ii s\hat\phi_i) \exp\left[-{\beta\over4}\left(F_n[\hat n_i+s]-F_n[\hat n_i]\right)\right], \qquad s = \pm 1.
\label{eq:secII_Lweak}
\eeqn
Since this jump operator does not commute with the total $\U(1)_c$ charge, the Lindbladian dynamics is only weakly symmetric.
We refer to this as Model A dynamics, because in Model A of Halperin-Hohenberg one has purely relaxation dynamics for a complex field, $\psi \sim e^{i\phi}$, with no conserved charge carrying field - that is, it is weakly symmetric dynamics.
In the Keldysh action the non-charge conserving jump operator leads to an additional term of the form,
\beqn
S_w = \gamma_w\int\dd t\sum_i (2\cos\tilde\phi_i -  \ii \beta K n_i\sin\tilde\phi_i).
\label{eq:secII_Sweak}
\eeqn
There is an explicit (strong) U(1)-breaking term $\cos \tilde{\phi}$ for the rotor degree of freedom $\tilde{\phi}$. The SW-SSB phase transition for $\tilde{\phi}$ will be washed out by this term, \cx{meaning $\tilde\phi$ is always in its spin-wave regime, or in other words with model-A dynamics (i.e. with weak symmetry), the Mott insulator and normal fluid are one and the same phase}. After expanding for small $\tilde{\phi}$ up to quadratic order, and decoupling by introducing noise fields, Eq.~\ref{eq:secII_Sweak} gives local relaxation of the conserved density,
\beqn
\partial_t n = -\Gamma_n n + \zeta_w, \qquad \Gamma_n = \beta\gamma_wK,
\eeqn
with standard noise
$\overline{\zeta_w(x,t)\zeta_w(x',t')}=2\gamma_w\delta(x-x')\delta(t-t')$.
The charge diffusion is suppressed by the explicit breaking of the strong $\U(1)_c$ symmetry. 

In the superfluid phase, the long-wavelength equations become
\beqn
\partial_t\phi &=& u_n n + D_\phi\nabla^2\phi + \eta, \cr\cr \partial_t n &=& u_\phi\nabla^2\phi + D_n\nabla^2 n - \Gamma_n n + \nabla\cdot\zeta + \zeta_w.
\label{eq:secII_modelA_H}
\eeqn
Since the charge density quickly relaxes to zero, the only remaining hydrodynamic mode in the superfluid is phase diffusion,
\beqn
\partial_t\phi = D_{\phi,{\rm eff}}\nabla^2\phi + \eta_{\rm eff},
\qquad
D_{\phi,{\rm eff}} = D_\phi +{ u_nu_\phi\over\Gamma_n}.
\label{eq:secII_modelA_phi_diffusion}
\eeqn
\uline{In the dual vortex formulation that we discuss in the next section, this diffusive mode corresponds to the diffusion of the dual electric field $E_T$. We will see that there is duality between the 0-form $\U(1)_c$ WT-SSB (model-A transition into the superfluid) and a 1-form $\U(1)^{(1)}_e$ SW-SSB (model-F). }

\begin{figure}
    \centering
    \includegraphics[width=0.7\linewidth]{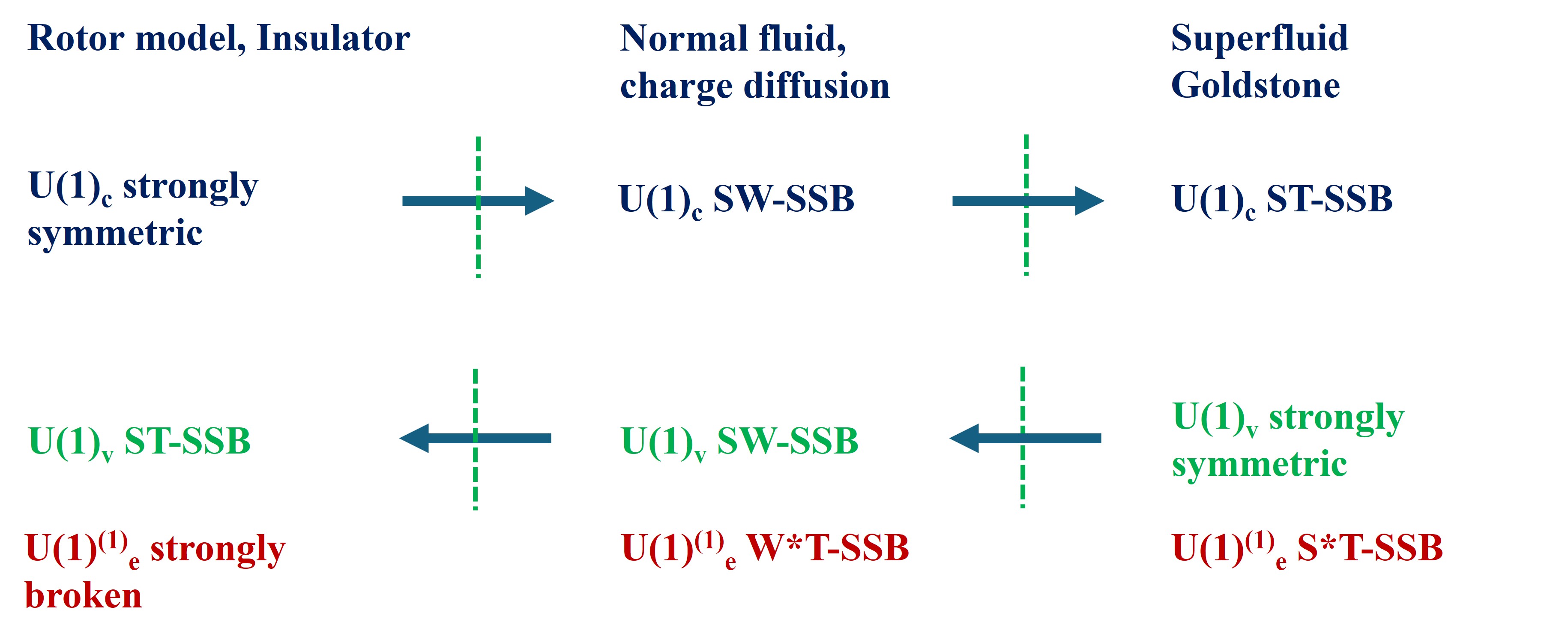}
    
    \caption{The dynamical phase diagram of quantum rotor model with $\U(1)_c$ symmetry. Starting with the insulator phase, after $\U(1)_c$-SW-SSB (condensing $\tilde{\phi}$), the system enters the normal fluid phase with charge density diffusion; after $\U(1)_c$-ST-SSB (condensing $\phi$), the system enters the superfluid phase with Goldstone mode. The same phase diagram can also be produced starting with the superfluid phase, which is the strongly symmetry phase of $\U(1)_v$. After the $\U(1)_v$-SW-SSB (condensing $\tilde{\theta}$), the system enters the normal fluid, and after $\U(1)_v$-ST-SSB (condensing $\theta$), the system enters the insulator phase of rotor. }
    
    \label{duality}
\end{figure}

\section{Particle-vortex duality of 2d hydrodynamics}

\label{2dPVduality}

The objective of this section is to develop a particle-vortex dual formalism for our rotor Lindbladian dynamics, and its emergent classical hydrodynamics.
The particle-vortex duality for bosonic systems in equilibrium was discussed in Ref.~\cite{halperindual,leefisherdual}. We will first consider the case with strong-U(1)$_c$ (model-F), then include jump operators that explicitly break strong-U(1)$_c$ to weak-U(1)$_c$ (model-A).

We consider gauge field $A_\ell$ on links of the dual two-dimensional square lattice, with conjugate electric field $E_\ell$, and matter rotors $\theta_i$ on sites, with conjugate density $N_i$:
\beqn
[\hat{A}_\ell,\hat{E}_{\ell'}]=\ii\delta_{\ell\ell'},
\qquad
[\hat{\theta}_i,\hat{N}_j]=\ii\delta_{ij}.
\eeqn
Here $\theta_i$ and $N_i$ are phase and density fields of the dual vortex.
The electric field and gauge field take on real values.  The gauge field is non-compact, so that there is no monopole in the $(2+1)d$ spacetime. 
For an oriented link $\ell=(i\rightarrow j)$, we use $A_{ji}=-A_{ij}$, $E_{ji}=-E_{ij}$. We also define magnetic flux through a plaquette $p$ 
\beqn
B_p=(\nabla\times A)_p = \sum_{\ell\in\partial p}s_{p\ell}A_\ell .
\eeqn
The quantum rotor Hamiltonian can be expressed in terms of the dual vortex fields as,
\begin{equation}
\hat{H} = \frac{u_E}{2} \sum_\ell \hat{E}_\ell^2 + \frac{u_B}{2} \sum_p \hat{B}_p^2 - J_v \sum_{\langle ij \rangle} \cos(\hat{\theta}_i - \hat{\theta}_j - \hat{A}_{ij}),
\label{eq:H_vortex}
\end{equation}
where $J_v$ is the strength of the vortex hopping.  This Hamilonian must be augmented by the gauge contraint that $\nabla \cdot \hat{E} = \hat{N}$.

\cx{Let us first discuss the symmetries of the dual vortex theory in equilibrium. There are three symmetries in total: 

    




\begin{itemize}

\item The original global $\U(1)_c$ symmetry becomes a topological magnetic-flux symmetry in the dual description. Its conserved current is
\beqn
j_{c,\mu}={1\over 2\pi}\epsilon_{\mu\nu\rho}\partial_\nu A_\rho,
\eeqn
so that the original $\U(1)_c$ charge is identified with the magnetic flux,
\beqn
Q_c = {1\over 2\pi}\int \dd^2x B.
\eeqn
The conservation of $Q_c$ follows from the absence of monopole operators for the noncompact dual gauge field.

\item One may also introduce an auxiliary ``vortex-number symmetry'', which we denote by $\U(1)_v$. The vortex field is electrically charged under the dual gauge field, and its phase rotation is therefore part of the gauge group. Nevertheless, when gauge fluctuations are neglected, the phase rotation of the vortex field may be treated as an effective $\U(1)_v$ symmetry. 

\item In the superfluid phase, vortices are gapped. At energies well below the vortex gap, dynamical electric charges of the dual gauge field (which are the vortices) are therefore absent, and an emergent electric $1$-form symmetry $\U(1)^{(1)}_e$ appears. Its conservation law is the source-free Maxwell equation $\partial_\mu F_{\mu\nu}=0$, which implies the conservation of the dual electric flux through a one-dimensional line.

In the superfluid regime, the electric field is related to the rotor current through
\beqn
E_i \sim \rho_s \epsilon_{ij}\partial_j\phi.
\eeqn
Consequently, the dual electric flux through a line measures the phase winding along that line. For example, the electric flux through a line normal to the $y$ direction is proportional to
\beqn
\int \dd x\,\partial_x\phi,
\eeqn
namely the winding number of the superfluid phase around the corresponding spatial cycle. We stress again that $\U(1)^{(1)}_e$ should be understood as an emergent infrared symmetry rather than an exact microscopic symmetry.

\end{itemize}
}

\subsection{Particle-vortex dual of Model-F hydrodynamics}


We first discuss three classes of Lindblad jump operators, all of which preserve the strong $\U(1)_c$ symmetry, which in this dual description corresponds to the conservation of magnetic flux $B$.  The targeted free energy is taken to be
\beqn
F[N,A,E] = {K\over2}\sum_\ell E_\ell^2 + {K_v\over2}\sum_i N_i^2 +
{J\over2}\sum_p B_p^2 .
\label{eq:FquadraticE}
\eeqn 
\cx{For generality, we have chosen a different convention of parameters in Eq.~\ref{eq:FquadraticE}, from the literal dual of the free energy of the superfluid phase Eq.~\ref{freeenergyrotorSF}, which would take the form $F \sim \int d^2x \frac{J_R}{2} E^2 + \frac{K}{2} B^2 $. }

{\it --- Gauge-invariant vortex hopping}

The gauge-invariant vortex hopping jump operator is taken to be,
\beqn
\hat L^{s}_{v,ij} = \sqrt{\gamma_v} \exp\left[ \ii s(\hat\theta_i-\hat\theta_j-\hat A_{ij}) \right] \exp\left[ - {\beta\over4}\Delta F^s_{v,ij} \right], \qquad s=\pm1 .
\label{eq:LvquadraticE}
\eeqn
To leading order in the change of free energy,
\beqn
\Delta F^s_{v,ij} \simeq s\left[ K_v(\hat N_i-\hat N_j)-K\hat E_{ij} \right].
\eeqn
Thus
\beqn
\hat L^{s}_{v,ij} \simeq \sqrt{\gamma_v} \exp\left[ \ii s(\hat\theta_i-\hat\theta_j-\hat A_{ij}) \right] \exp\left[ - {\beta s\over4} \left( K_v(\hat N_i-\hat N_j)-K\hat E_{ij} \right) \right].
\eeqn
The corresponding contribution to the Lindbladian is,
\begin{equation}
 {\cal L}_v [\hat{\rho}] = \sum_{\langle i, j \rangle} \sum_{\pm}  [ \hat{L}^\pm_{v_{ij}}\hat{\rho}  \hat{L}^{\pm \dagger}_{v_{ij}} - \frac{1}{2} \{ \hat{L}^{\pm \dagger}_{v_{ij}} \hat{L}^\pm_{v_{ij}} , \hat{\rho}\} ].
\end{equation}

{\it --- Link-$A$ jump}

The link-$A$ jump shifts $A_\ell$:
\beqn
\hat L^a_{A,\ell} = \sqrt{\gamma_A} \exp(-\ii a\hat E_\ell) \exp\left[ - {\beta\over4} \left( F[\hat A_\ell+a]-F[\hat A_\ell] \right) \right], \qquad a\in\mathbb R
\label{eq:LAquadraticE}
\eeqn
with $a$ taken as infinitesimal, so that the corresponding term in the Lindbladian is,
\begin{equation}
 {\cal L}_A[\hat{\rho}] =  \sum_\ell \int_a \partial_a^2 \delta(a) [ \hat{L}^a_{A,\ell}\hat{\rho}  \hat{L}^{a \dagger}_{A,\ell} - \frac{1}{2} \{ \hat{L}^{a \dagger}_{A,\ell} \hat{L}^a_{A,\ell} , \hat{\rho}\} ].
 \end{equation}
Since
\beqn
{\delta F\over\delta A_\ell} = J\sum_{p\supset\ell}s_{p\ell}B_p ,
\eeqn
this jump generates diffusion of the magnetic flux $B$.

{\it --- Closed-loop electric-field jump}

The closed-loop electric-field jump is
\beqn
\hat L^a_{E,p} = \sqrt{\gamma_E} \exp(\ii a\hat B_p) \exp\left[ -{\beta\over4} \left( F[\hat E+a\,\partial p]-F[\hat E] \right) \right], \qquad a\in\mathbb R .
\label{eq:LEquadraticE}
\eeqn
We note that here $a$ takes continuous values in $\mathbb R$ rather than discrete values because the gauge field is noncompact. In the appendix when we discuss the dynamics of compact gauge field, $a$ will be restricted to discrete values. The corresponding term in the Lindbladian is,
\begin{equation}
 {\cal L}_E[\hat{\rho}] =  \sum_p \int_a \partial_a^2 \delta(a) [ \hat{L}^a_{E,p}\hat{\rho}  \hat{L}^{a \dagger}_{E,p} - \frac{1}{2} \{ \hat{L}^{a \dagger}_{E,p} \hat{L}^a_{E,p} , \hat{\rho}\} ].
 \end{equation}
The corresponding detailed-balance force is
\beqn
\sum_{\ell\in\partial p}s_{p\ell}{\delta F\over\delta E_\ell} = K\sum_{\ell\in\partial p}s_{p\ell}E_\ell = K(\nabla\times E)_p .
\eeqn
Therefore this jump generates diffusion of the transverse electric field.

The full Lindblad equation of motion is taken as,
\begin{equation}
\partial_t \hat{\rho} = - \ii [ \hat{H}, \hat{\rho} ] + {\cal L}_v [\hat{\rho}] + {\cal L}_A [\hat{\rho}] + {\cal L}_E[\hat{\rho}].
\end{equation}

Once again we re-express the Lindbladian dynamics in terms of a Keldysh path integral, but now over the fields $A_\ell$, $E_\ell$, $N_i$ and $\theta_i$ for both forward and backward paths (denoted $R/L$).  The kinematic part of the Keldysh action, including Gauss law, is
\beqn
S_0 + S_G = \ii\int\dd t \sum_\ell \left( \tilde E_\ell\dot A_\ell + E_\ell\dot{\tilde A}_\ell \right) + \sum_i \left( \tilde N_i\dot\theta_i + N_i\dot{\tilde\theta}_i \right) \cr\cr - \ii\int\dd t \sum_i \tilde A_{0,i} \left( \nabla\cdot E - N \right)_i + A_{0,i} \left( \nabla\cdot\tilde E-\tilde N \right)_i.
\label{eq:S0SG_quadratic}
\eeqn
As with the rotors, we have defined classical (un-tilded) and quantum (tilded) fields, also for the scaler potential that enforces Gauss' law,  
\beqn
A_{0,i}={A_{0,R,i} + A_{0,L,i}\over2}, \qquad \tilde A_{0,i} = A_{0,R,i}-A_{0,L,i}.
\eeqn

The vortex-hopping jump operator gives the following contribution to the Keldysh action,
\beqn
S_v = \gamma_v\int\dd t \sum_{\langle i,j\rangle} 2\cos(\tilde\theta_i-\tilde\theta_j-\tilde A_{ij}) - \ii\beta \left( K_v(N_i-N_j)-K E_{ij} \right) \sin(\tilde\theta_i-\tilde\theta_j-\tilde A_{ij}).
\label{eq:SvquadraticE}
\eeqn
Similarly, the link-$A$ jump gives
\beqn
S_A = \gamma_A\int\dd t \sum_\ell - \tilde E_\ell^2 + \ii\beta J\tilde E_\ell \sum_{p\supset\ell}s_{p\ell}B_p,
\label{eq:SAquadraticE}
\eeqn
and the closed-loop electric-field jump gives,
\beqn
S_E = \gamma_E\int\dd t \sum_p - \tilde B_p^2 - \ii\beta K(\nabla\times E)_p \tilde B_p.
\label{eq:SEquadraticE}
\eeqn
The full Keldysh action is,
\beqn
S=S_0+S_G+S_v+S_A+S_E .
\eeqn
We will also consider an additional coherent Hamiltonian dynamics term below.

\subsubsection{Superfluid phase} \label{dualsuper}

We first discuss the standard superfluid phase of $\U(1)_c$. In the dual gauge-field language, the vortices are gapped, and we should not perform a polynomial-expansion of $\tilde Q_{ij}$, where 
\beqn
\tilde Q_{ij}=\tilde\theta_i-\tilde\theta_j-\tilde A_{ij}.
\eeqn
Equivalently, the vortex density $N$ is not a hydrodynamic variable. The Gauss-law constraint then ties the non-hydrodynamic longitudinal electric field to the non-hydrodynamic vortex density: $\nabla\cdot E_L=N$. The hydrodynamic variables are instead the transverse electric field $\vE_T$ and the magnetic flux $B$.

If we first ignore coherent Hamiltonian dynamics and set $\gamma_v=0$, we can decouple the $\tilde{E}_\ell^2$ and $\tilde{B}_p^2$ terms by introducing noise terms, and then - upon integrating over $\tilde{E}_\ell$ and $\tilde{A}_\ell$ we get classical hydrodynamic equations of motion for the classical $B$ and $E_T$ fields,
\beqn
\partial_t\vE_T &=& D_E\nabla^2\vE_T -\hat z\times\nabla\eta^E, \cr\cr \partial_tB &=& D_B\nabla^2B + \nabla\times\vect{\xi}^{A}.
\label{eq:case11_quadratic}
\eeqn
Here, we have defined
\beqn
D_E = \beta\gamma_EK, \qquad D_B = \beta\gamma_AJ ,
\eeqn
and taken the spatial continuum limit.
The noise fields satisfy $
\overline{\xi^A_\ell(t)\xi^A_{\ell'}(t')} = 2\gamma_A\delta_{\ell\ell'}\delta(t-t')$, $\overline{\eta^E_p(t)\eta^E_{p'}(t')} = 2\gamma_E\delta_{pp'}\delta(t-t')$.
Thus, without Hamiltonian dynamics, the superfluid has two diffusive hydrodynamic modes $E^T$ and $B$.
\uline{This is the dual-gauge-field version of the previous section that, in the incoherent superfluid, both $\phi$ and $n$ diffuse.}

Now let us include the coherent Maxwell Hamiltonian, Eq.~\ref{eq:H_vortex}, setting $J_v=0$ appropriate in the superfluid phase because vortices remain gapped. 
Hence $N$ and $E_L$ remain non-hydrodynamic. After adding the coherent Hamiltonian dynamics into the Keldysh action, the transverse sector now obeys,
\beqn
\partial_t\vE_T &=& D_E\nabla^2\vE_T + u_B\hat z\times\nabla B -\hat z\times\nabla\eta^E, \cr\cr
\partial_t B &=& u_E\nabla\times\vE_T + D_B\nabla^2B + \nabla\times\vect{\xi}^{A}.
\label{eq:B_case21_quadratic}
\eeqn
The temporal gauge field $A_0$ enforces Gauss law, but it drops out of the transverse equation because $\nabla\times\nabla A_0 = 0 $.
At small momentum, we obtain the mode
\beqn
\omega_\pm(k) = \pm ck - \ii {D_E+D_B\over2}k^2 + O(k^3).
\label{modelFHsuper}
\eeqn
Thus the Hamiltonian turns the two diffusive superfluid modes into a weakly damped propagating photon mode, which is the Goldstone mode of the dual 1-form symmetry~\cite{Gaiotto_2015,Hofman_2019}. \uline{The photon is the dual description of the rotor Goldstone mode, as was well-established in duality of equilibrium systems}

The superfluid is a phase where the vortices are massive, therefore it is a strongly symmetric phase of $\U(1)_v$. There is also a dual emergent-strong-1-form symmetry $\U(1)^{(1)}_e$ associated with the conservation of electric flux $\vE$ in the infrared. In the superfluid phase this dual emergent $\U(1)^{(1)}_e$ is spontaneously broken to its trivial subgroup, which we label as S$^\ast$T-SSB, where ``$^\ast$" stands for emergent symmetry.

\subsubsection{Normal fluid phase} \label{dualnormal}

We now discuss the normal fluid phase.  On the rotor side this is the phase with SW-SSB of $\U(1)_c$, but without spontaneous breaking of the weak $\U(1)_c$ symmetry.  Therefore the charge density $n$ is hydrodynamic, but the phase $\phi$ is not.  Under particle-vortex duality,
$ n \sim B$, and $\hat z\times\nabla\phi\leftrightarrow \vE_T$, so the dual statement is that $B$ remains hydrodynamic, while $\vE_T$ is not hydrodynamic.

In the dual formalism, the normal fluid most naturally corresponds to the condensate of $\tilde{\theta}$, i.e. the SW-SSB of $\U(1)_v$, therefore we can perform a polynomial-expansion of $\tilde Q_{ij}$. After introducing a decoupling noise field, we arrive at a classical equation of motion for the classical electric field,
\beqn
\dot E_{ij} = \beta\gamma_v \left[ K_v(N_j-N_i)-K E_{ij} \right] + \eta^v_{ij} + \sum_{p\supset ij}s_{p,ij}\eta^E_p +\cdots .
\label{eq:Edot_quadratic_noH_lattice}
\eeqn
The vortex density equation is fixed by Gauss law
$ \dot N_i=(\nabla\cdot\dot E)_i$, and hence it is not an independent mode. 

The magnetic flux obeys,
\beqn
\dot B_p = -\beta\gamma_AJ \sum_{\ell\in\partial p}s_{p\ell} \sum_{p'\supset\ell}s_{p'\ell}B_{p'} + \sum_{\ell\in\partial p}s_{p\ell}\xi^A_\ell.
\label{eq:Bdot_quadratic_noH_lattice}
\eeqn 
It is convenient to define,
\beqn
D_N=\beta\gamma_vK_v, \qquad \Gamma_E=\beta\gamma_vK, \qquad D_E=\beta\gamma_EK, \qquad D_B=\beta\gamma_AJ.
\eeqn
Considering the spatial continuum limit, 
Eq.~\ref{eq:Bdot_quadratic_noH_lattice} becomes the diffusion of magnetic flux,
\beqn
\partial_tB = D_B\nabla^2B + \nabla\times\vect{\xi}^{A}.
\label{eq:B_quadratic_noH}
\eeqn

The continuum limit equation of motion of the electric field reads,
\beqn
\partial_t\vE = D_N\nabla N - \Gamma_E\vE + D_E\nabla^2\vE_T + \vect{\eta}^{v} -\hat z\times\nabla\eta^E +\cdots.
\label{eq:E_quadratic_noH}
\eeqn
Using Gauss law $N = \nabla\cdot E$,
we obtain
\beqn
\partial_tN
= D_N\nabla^2N - \Gamma_E N + \nabla\cdot\vect{\eta}^{v},
\label{eq:N_quadratic_noH}
\eeqn
namely the vortex density $N$ is locally relaxed. 
Therefore the only hydrodynamic mode in the normal fluid is magnetic-flux diffusion. 
\uline{Thus the condensate of $\tilde{\theta}$ gives a dual description of the normal fluid phase of the rotor model, with $n$ a hydrodynamic field but $\phi$ is not.} 

Let us also take the perspective of $\U(1)^{(1)}_e$. As we mentioned before, in the superfluid phase there is a dual emergent-strong-$\U(1)^{(1)}_e$ symmetry, since all vortices are massive. In the normal fluid phase, since $\tilde{\theta}$ condenses (there is a $\U(1)_v$ SW-SSB), the emergent-strong-$\U(1)^{(1)}_e$ symmetry is broken, but there is still an emergent-weak-$\U(1)^{(1)}_e$ symmetry. The spontaneous breaking of this emergent-weak-$\U(1)^{(1)}_e$ symmetry leads to hydrodynamic diffusion mode of $B$. In the appendix we will see that if one starts with a lattice gauge theory with explicit 1-form symmetry, the WT-SSB of the $\U(1)^{(1)}_e$ will indeed lead to diffusion of $B$. \uline{Therefore the SW-SSB (model-F) of $\U(1)_c$ is dual to the W$^\ast$T-SSB (model-A) of $\U(1)^{(1)}_e$. } The diffusion of magnetic flux $B$ is a key ingredient of magnetohydrodynamics. The connection between magnetohydrodynamics and 1-form symmetry has been explored in recent years~\cite{magneto1,magneto2,magneto3,magneto4,magneto4,magneto5,magneto6,magneto7,Yoshimura_2026}.

It may at first seem surprising that the vortex density does not exhibit an independent diffusive mode in the normal phase, even though vorticity is locally conserved. The reason is that the vortices are gauge charges and are therefore tied to the longitudinal electric field through Gauss law, $N=\nabla\cdot E$. Eliminating the longitudinal electric field gives the effective quadratic free energy
\beqn
F_{\rm eff}[N] = \frac{1}{2}\int_{\bm k} \left( K_v+\frac{K}{k^2} \right) |N(\bm k)|^2 .
\eeqn
The conserved vortex-hopping dynamics has mobility proportional to \(k^2\), and hence
\beqn
\partial_t N(\bm k)
=
-\beta\gamma_v k^2
\frac{\delta F_{\rm eff}}{\delta N(-\bm k)}
+\ii\bm k\cdot\bm\eta^v
=
-\beta\gamma_v
\left(
K+K_vk^2
\right)
N(\bm k)
+\ii\bm k\cdot\bm\eta^v .
\eeqn
Thus the \(1/k^2\) Coulomb interaction generated by the longitudinal electric field cancels the \(k^2\) factor associated with conserved hopping, producing a finite relaxation rate $\Gamma_N = \Gamma_E = \beta\gamma_v K$ as \(k\rightarrow0\). The vortex density therefore undergoes relaxation rather than ordinary diffusion: its long-wavelength mode is gapped, with only a subleading diffusive correction \(D_Nk^2\). Consequently, the magnetic flux \(B\), rather than the vortex density \(N\), is the only hydrodynamic diffusive variable in the normal phase.


\subsubsection{Mott insulator phase} \label{dualMott}

Finally we briefly discuss the Mott insulator phase. On the rotor side, this is the phase before SW-SSB of the strong $\U(1)_c$ symmetry. The microscopic discreteness of the charge density $n$ remains visible, so $n$ should not yet be treated as a continuous hydrodynamic field. The Mott insulator is the phase where all vortices condense, hence it is a phase where the dual emergent $\U(1)^{(1)}_e$ is strongly broken, and the $\U(1)_v$ has ST-SSB. The phase diagram of the rotor model with strong $\U(1)_c$ symmetry can therefore be produced with both the original rotor formalism, and the dual vortex formalism in this section, as illustrated in Fig.~\ref{duality}. 

Here we note that there is a mixed perturbative 't Hooft anomaly between $\U(1)_c$ and $\U(1)^{(1)}_e$, hence there is no short-range entangled phase where both symmetries are strongly symmetric, namely $\U(1)_c$ and $\U(1)^{(1)}_e$ cannot both be in their insulating phase~\cite{superfluidanomaly1,superfluidanomaly2}. Such phase would violate the 't Hooft anomaly matching condition.

\subsection{Model-A hydrodynamics}

In this section we consider Lindblad jump operators with weak-$\U(1)_c$ symmetry, which we again refer to as Model A. We add a Lindblad jump operator that locally creates or annihilates magnetic flux:
\beqn
\hat L^s_{B,p} = \sqrt{\gamma_B}  \exp(\ii s\hat\phi_p) \exp\left[ -{\beta\over4}\Delta F^s_{B,p} \right], \qquad s=\pm1,
\eeqn
where $[\hat{\phi}_p,\hat{B}_{p'}]=\ii\delta_{pp'}.$In the dual picture this jump operator is also the monopole operator of the gauge field $A_\ell$. 

The corresponding term in the Keldysh action is,
\beqn
S_B = \gamma_B\int\dd t \sum_p 2\cos\tilde\phi_p - \ii\beta JB_p\sin\tilde\phi_p.
\eeqn
Here $\cos(\tilde\phi_p)$ is the monopole creation/destruction term, which is known to be strongly relevant and keeps $\vect{A}$ in its confined phase, or equivalently \uline{the weak-$\U(1)^{(1)}_e$ symmetry must remain symmetric and unbroken~\cite{Gaiotto_2015}}. Expanding in $\tilde\phi_p$ and introducing a noise decoupling gives local relaxation of the classical magnetic field $B$:
\beqn
\partial_t B = -\Gamma_B B + \zeta_B, \qquad \Gamma_B = \beta \gamma_B J,
\eeqn
with $\overline{\zeta_B(p,t)\zeta_B(p',t')} = 2\gamma_B\delta_{pp'}\delta(t-t')$.

In the superfluid phase,  $\nabla\tilde\theta-\tilde A$ is not condensed, thus $N$ is not hydrodynamic. The equations for the transverse sector of the gauge fields become,
\beqn
\partial_t\vE_T &=& D_E\nabla^2\vE_T + u_B\hat z\times\nabla B -\hat z\times\nabla\eta^E,
\cr\cr
\partial_tB &=& u_E\nabla\times\vE_T + D_B\nabla^2B - \Gamma_BB + \nabla\times\vect{\xi}^{A} + \zeta_B .
\label{eq:case31_quadratic}
\eeqn
For $\Gamma_B\neq0$ and small $k$,
we obtain modes
\beqn
\omega_{\rm slow}(k) &=& -\ii \left( D_E+{u_Eu_B\over\Gamma_B} \right)k^2 + O(k^4), \cr\cr
\omega_{\rm fast}(k) &=& -\ii\Gamma_B - \ii \left( D_B - {u_Eu_B\over\Gamma_B} \right)k^2 + O(k^4).
\eeqn
Thus the single-flux jump (the monopole creation/annihilation) operator destroys the propagating transverse photon mode in the infrared and leaves a diffusive $E_T$ mode.

The superfluid phase again has a dual emergent-strong-$\U(1)^{(1)}_e$ symmetry. In the perspective of the dual $\U(1)^{(1)}_e$ symmetry, the single flux jump operator corresponds to a term $\cos(\tilde{\phi})$ in the Keldysh action, which is the monopole creation/destruction term of the gauge flux $B$. The common wisdom is that the monopole term will keep the gauge field $\vA$ in its confined phase, which is also the symmetric phase of weak-$\U(1)^{(1)}_e$. 
Therefore the emergent-strong-$\U(1)^{(1)}_e$ symmetry can only have S$^\ast$W-SSB, rather than S$^\ast$T-SSB. \uline{ The diffusion of $\vE_T \sim \hat{z} \times \nabla \phi$ can be viewed as the consequence of WT-SSB of $\U(1)_c$, or equivalently the S$^\ast$W-SSB of the dual $\U(1)^{(1)}_e$}. 
In the appendix we will also show that if one starts with a $2d$ compact gauge theory with an explicit strong $\U(1)^{(1)}_e$ symmetry, the SW-SSB of $\U(1)^{(1)}_e$ does lead to diffusion of $\vE$. Therefore \uline{the WT-SSB of $\U(1)_c$ (model-A) is dual to the S$^\ast$W-SSB of $\U(1)^{(1)}_e$ (model-F)}.

In the normal fluid, $\nabla\tilde\theta-\tilde A$ is condensed. Including both the quadratic electric-field relaxation and the single-flux relaxation, the continuum equations become
\beqn
\partial_t\vE &=& D_N\nabla N - \Gamma_E\vE + D_E\nabla^2\vE_T + u_B\hat z\times\nabla B + \vect{\eta}^{v} -\hat z\times\nabla\eta^E +\cdots ,
\cr\cr
\partial_tB &=& u_E\nabla\times\vE + D_B\nabla^2B - \Gamma_B B + \nabla\times\vect{\xi}^{A} + \zeta_B .
\label{eq:case32_quadratic}
\eeqn

The vortex density $N$ is again tied with the longitudinal mode $E_L$: 
$ N=\nabla\cdot E$.
Therefore
\beqn
\partial_tN = D_N\nabla^2N - \Gamma_E N + \nabla\cdot\vect{\eta}^{v}.
\eeqn
Thus $N$ is locally relaxed and is not hydrodynamic.

For $\Gamma_E\neq0$ and $\Gamma_B\neq0$, both $B$ and transverse $E_T$ are locally relaxed. 
Therefore there is no transverse hydrodynamic mode in the normal fluid, consistent with the model-A dynamics discussed in the previous section. Since $\tilde{\theta}$ is condensed, the $\U(1)_v$ has SW-SSB in the normal fluid, and there is no emergent-strong-$\U(1)^{(1)}_e$ symmetry. But since $\theta$ is not condensed, there is still an emergent-weak-$\U(1)^{(1)}_e$ symmetry, but due to the $\cos(\tilde{\phi})$ term in the action, this emergent-weak-$\U(1)^{(1)}_e$ symmetry is not spontaneously broken, i.e. the system remains weakly$^\ast$-symmetric. \uline{Therefore the normal fluid of model-A is dual to the weakly$^\ast$-symmetric phase of $\U(1)^{(1)}_e$, consistent with the fact that there is no hydrodynamics mode. }

\section{Summary and Outlook}

In this work, we explored the emergence of classical hydrodynamics from quantum Lindbladian formulations related by duality. Using particle-vortex duality for $2d$ boson and rotor systems, we showed that the same hydrodynamic phase diagram can be described either in terms of the charge symmetry $\U(1)_c$ of the decohered rotor model, or equivalently in terms of the vortex symmetry $\U(1)_v$ and the emergent dual one-form symmetry $\U(1)^{(1)}_e$. In particular, we found that the model-F/model-A dynamics associated with SW-SSB/WT-SSB of $\U(1)_c$ is dual to the model-A/model-F dynamics associated with WT-SSB/SW-SSB of $\U(1)^{(1)}_e$, respectively. In the Appendix, we further studied the decoherence dynamics of a lattice gauge theory with an exact $\U(1)^{(1)}_e$ symmetry, providing additional support for this duality. 

A parallel generalization of our work to $1d$ and $3d$ will lead to self-dual hydrodynamics. For example, in $1d$, the model-F dynamics from SW-SSB of a $\U(1)_c$ 0-form symmetry is dual to the model-A dynamics from W$^\ast$T-SSB of a dual emergent 0-form symmetry; and in $3d$ the model-F dynamics of a 1-form symmetry is dual to the model-A dynamics of a dual emergent 1-form symmetry. The general pattern we uncover is that, \uline{in $d$-dim space, the model-F hydrodynamics of a $n$-form symmetry as a consequence of SW-SSB, is dual to the model-A hydrodynamics of a $(d - n - 1)$-form symmetry, arising from WT-SSB}.

Our dual formalism treats current and vorticity on the same footing, which are the ingredients of the classic Navier-Stokes equation. In an upcoming work we will explore how the full Navier-Stokes  equation might emerge from a similar setting. 

Cenke Xu is supported by the Simons foundation through the Simons Investigator program. M.P.A.F. is supported by the
Simons Collaboration on Ultra-Quantum Matter, which is a grant from the Simons Foundation 651457, and also a Quantum Interactive Dynamics grant from the William M. Keck Foundation. We thank Sagar Vijay for helpful discussions.  We also acknowledge using generative AI for discussion and lengthy computation.

\appendix

\bibliography{hydro}

\section{$2d$ compact $\U(1)$ gauge theory} \label{2dcompactgauge}

\begin{table}[h]
\centering
\scriptsize
\renewcommand{\arraystretch}{2.0}
\begin{tabular}{|l|l|l|l|l|}
\hline
Dynamics & Phase & $\U(1)^{(1)}_e$ & $\U(1)_c$ & Hydro modes \\
\hline

Model-F, strong $\U(1)^{(1)}_e$
&
Confined phase
&
Strongly Symmetric
&
Strongly Broken
&
none
\\
\hline

Model-F, strong $\U(1)^{(1)}_e$
&
1-form normal fluid
&
SW-SSB
&
W$^\ast$T-SSB
&
diffusion of $\vE$
\\
\hline

Model-F, strong $\U(1)^{(1)}_e$
&
Photon phase
&
ST-SSB
&
S$^\ast$T-SSB
&
photon mode
\\
\hline

Model-A, weak $\U(1)^{(1)}_e$
&
1-form normal fluid
&
Weakly-Symmetric
&
Weakly$^\ast$-Symmetric
&
none
\\
\hline

Model-A, weak $\U(1)^{(1)}_e$
&
Magnetohydro phase
&
WT-SSB
&
S$^\ast$W-SSB
&
diffusion of $B$
\\
\hline

\end{tabular}
\caption{Symmetry structure and hydrodynamic modes for dynamics with strong or weak $\U(1)^{(1)}_e$ symmetry. Here $^\ast$ denotes an emergent symmetry.}
\label{tab:one_form_symmetry_hydro}
\end{table}

In the previous models, the 1-form symmetry was emergent. We now consider a compact \(\U(1)\) gauge theory in $2d$ with an exact electric 1-form symmetry (there is no explicit dynamical matter field). SW-SSB of discrete higher form symmetries were discussed before, see for instance Ref.~\cite{higherSWSSB}. In our current case we define compact \(A_\ell\sim A_\ell+2\pi\) and integer electric field \(E_\ell\) on links: \beqn [A_\ell,E_{\ell'}]=\ii\delta_{\ell\ell'}, \qquad B_p=(\nabla\times\vect A)_p . \eeqn We take the standard form of the free energy of compact lattice gauge theory: 
\beqn F[\vect E,\vect A] = \sum_\ell {K\over 2}E_\ell^2 - J\sum_p\cos B_p. \eeqn

The initial state of our dynamics is chosen to be the confined phase of the $\U(1)$ gauge field, it is almost an eigenstate of $\vect{E}_\ell = 0$, analogous to the charge Mott insulator considered in the main part of the paper. The kinetic term in the Keldysh action is \beqn S_B = \ii \int \dd t \sum_\ell ( \tilde{\vect{E}}_\ell\cdot  \dot{\vect{A}}_\ell + \vect{E}_\ell \cdot \dot{\tilde{\vect{A}}}_\ell ).  \eeqn

A $(2+1)d$ compact gauge field is dual to a rotor model, but there is no explicit 0-form $\U(1)_c$ symmetry of the dual rotor, because of the compactness of the gauge field. We can ``prepare" the initial confined phase by evolving the system in the imaginary time $t \in (-\infty, 0)$ with the ``generating" action of the confined phase. In particular, in terms of the dual rotor variables, the generating action for this state should include the following term \beqn S_v = \int_{-\infty}^0 dt \sum_r - v \cos(\phi_{Lr}) - v \cos(\phi_{Rr}) + \cdots. \eeqn The $\cos(\phi_{L,R})$ operator is the monopole creation/destruction operator of the compact gauge field, it also explicitly breaks the $\U(1)_c$ symmetry, and plays the role as the boundary-pinning effect for $\phi$ and $\tilde{\phi}$. Because of these boundary-pinning effects, the $\U(1)_c$ will need to ``emerge" at later time $t$. Again, due to the mixed perturbative 't Hooft anomaly between $\U(1)^{(1)}_e$ and $\U(1)_c$, a short-range entangled phase where both symmetries are strongly symmetric does not exist. 

We first consider jump operators that respect the strong electric 1-form symmetry. These jump operators include \beqn \hat L_{A,\ell}^a &=& \sqrt{\gamma_A} \exp(-\ii a\hat E_\ell) \exp\left[ -{\beta\over 4} \left( F_B[\hat A_\ell+a]-F_B[\hat A_\ell] \right) \right], \cr\cr \hat L_{E,p}^s &=& \sqrt{\gamma_E} \exp(\ii s\hat B_p) \exp\left[ -{\beta\over 4} \left( F_E[\vect E+s\,\partial p]-F_E[\vect E] \right) \right], \qquad s=\pm1, \eeqn where \beqn F_B[\vect A]=-J\sum_p\cos B_p, \qquad F_E[\vect E]=\sum_\ell {K\over 2}E_\ell^2 . \label{jump1-form} \eeqn We note here that $E_\ell$ only takes integer values. 
The corresponding terms in the Lindbladian are,
\begin{equation}
 {\cal L}_A[\hat{\rho}] =  \sum_\ell \int_a \partial_a^2 \delta(a) [ \hat{L}^a_{A,\ell}\hat{\rho}  \hat{L}^{a \dagger}_{A,\ell} - \frac{1}{2} \{ \hat{L}^{a \dagger}_{A,\ell} \hat{L}^a_{A,\ell} , \hat{\rho}\} ],
 \end{equation}
 and
 \begin{equation}
 {\cal L}_E[\hat{\rho}] =  \sum_p \sum_{\pm} [ \hat{L}^{\pm}_{E,p}\hat{\rho}  \hat{L}^{\pm \dagger}_{E,p} - \frac{1}{2} \{ \hat{L}^{\pm \dagger}_{E,p} \hat{L}^\pm_{E,p} , \hat{\rho}\} ].
 \end{equation}

To lowest nontrivial order in \(\beta\), ${\cal L}_A$ and ${\cal L}_E$ generate the following terms in the Keldysh action, \beqn S_A &=& \gamma_A\int\dd t\sum_\ell -\tilde E_\ell^2 + \ii\beta J\tilde E_\ell \sum_{p\supset\ell}s_{p\ell} (\sin B_{R,p}+\sin B_{L,p}) , \cr\cr S_E &=& \gamma_E\int\dd t\sum_p 2\cos\tilde B_p - \ii\beta K(\nabla\times\vect E)_p\sin\tilde B_p. \label{SASE} \eeqn

\subsection{SW-SSB of $\U(1)^{(1)}_e$} \label{2dcompactgaugeSWSSB}

\cx{We first clarify that for a compact gauge field in $2d$, there likely will not be a finite-time SW-SSB transition for $\U(1)^{(1)}_e$, meaning we cannot perform polynomial-expansion of $\tilde{B}$ in Eq.~\ref{SASE} at any finite time, in the rigorous infrared limit. Mathematically this is because a finite-time transition can often be mapped to a $2d$ classical-statistical mechanics model with $\U(1)^{(1)}_e$ symmetry, with time $t$ identified as $1/T$. But we know that there will not be any transition of such model in $2d$ as the compact gauge field is always in its confined phase. Another way of interpreting this is that, the boundary pinning effect from the monopole terms in $t \in (- \infty, 0)$ is strong and relevant. But we can still discuss the possibility of this expansion within finite length scale. For example, if $\gamma_E \gg \gamma_A$, and $\gamma_E t \gg 1$, $\tilde{B}$ fluctuates slowly over a large but finite SW-SSB length scale $\xi$, which justifies the expansion. } After expanding flux \(\tilde B=\nabla\times\tilde{\vect A}\), $S_E$ becomes \beqn
S_E = \gamma_E\int\dd t\sum_p -\tilde B_p^2
-\ii\beta K(\nabla\times\vect E)_p\tilde B_p.
\eeqn
Decoupling the $\tilde{B}_p^2$ term by introducing a noise field,
leads to diffusion of the classical $\vE$ field, again within the finite SW-SSB length scale:
\beqn
\dot{\vect E} = D_E\nabla^2\vect E + \hat{z} \times \nabla \eta, \quad \overline{\eta(t,\vect x)\eta(t',\vect x')}
=
2\gamma_E
\delta(t-t')\delta^{(2)}(\vect x-\vect x') .
\label{diff2dgaugeSW} \eeqn
Here, we have used the fact that there is no matter fields present, so the gauge constraint $\nabla \cdot \vect{E}=0$ is rigorously imposed.

\cx{Since we assumed that $\tilde{B}$ can be expanded within scale $\xi$, within this length scale the compactness of the gauge field $\tilde{A}$ may be ignored, and there is an emergent-dual-0-form symmetry $\U(1)_c$, associated with the conservation of $\tilde{B}$.}  This is a weak-$\U(1)_c$, and it is spontaneously broken within scale $\xi$. \uline{Eq.~\ref{diff2dgaugeSW} is the same as Eq.~\ref{eq:case31_quadratic} after ignoring the relaxing $B$, and also the dual of Eq.~\ref{eq:secII_modelA_phi_diffusion}, arising from W$^\ast$T-SSB of $\U(1)_c$. Therefore in $2d$ there is a duality between 1-form model-F and 0-form model-A}. 

\subsection{ST-SSB of $\U(1)^{(1)}_e$}

We now consider the phase with ST-SSB of the electric 1-form symmetry (again within certain finite length scale $\xi$), which is often called the photon phase of the gauge field. In this phase the compactness of the gauge fields may be ignored. Equivalently, monopoles remain perturbative within this scale, so the dual $\U(1)_c$ emerges as a strong symmetry.
In this phase, both $\tilde B$ and $B$ can be expanded. 
Therefore, without coherent Hamiltonian dynamics, the leading hydrodynamic equations are
\beqn
\partial_t\vE &=& D_E\nabla^2\vE + \hat z\times\nabla\eta^E, \cr\cr \partial_tB &=& D_B\nabla^2B + \nabla\times\vect{\xi}^{A},
\label{eq:photonNoHcompact}
\eeqn
where
\beqn
D_E = \beta\gamma_EK, \qquad D_B = \beta\gamma_AJ .
\eeqn
The noises satisfy
\beqn
\overline{\eta^E(t,\vect x)\eta^E(t',\vect x')} = 2\gamma_E\delta(t-t')\delta^{(2)}(\vect x-\vect x'), \quad \overline{\xi^A_\ell(t)\xi^A_{\ell'}(t')} = 2\gamma_A\delta_{\ell\ell'}\delta(t-t') .
\eeqn

If we include the coherent Maxwell Hamiltonian
\beqn
H={u_E\over2}\sum_\ell E_\ell^2+{u_B\over2}\sum_pB_p^2 ,
\label{eq:compactPhotonH}
\eeqn
then the hydrodynamic equations become
\beqn
\partial_t\vE &=& D_E\nabla^2\vE + u_B\hat z\times\nabla B + \hat z\times\nabla\eta^E,
\cr\cr
\partial_tB &=& u_E\nabla\times\vE + D_B\nabla^2B + \nabla\times\vect{\xi}^{A}.
\label{eq:photonHcompact}
\eeqn
At small momentum,
\beqn
\omega_\pm(k) = \pm ck - \ii {D_E+D_B\over2}k^2 + O(k^3),
\qquad
c=\sqrt{u_Eu_B}.
\label{eq:compactPhotonMode}
\eeqn
Thus the ST-SSB phase of the exact electric 1-form symmetry supports a weakly damped photon mode. \uline{This photon mode is precisely the dual of the Goldstone mode arising from the ST-SSB of the $\U(1)_c$. }

\subsection{Weak $\U(1)^{(1)}_e$ symmetry and its WT-SSB} \label{2dcompactgaugeWTSSB}

Now we preserve only the weak-$\U(1)^{(1)}_e$ symmetry by allowing the open-string jumps \beqn \hat J_{E,\ell}^s = \sqrt{\gamma_w} \exp(\ii s\hat A_\ell) \exp\left[ -{\beta\over 4} \left( F_E[\hat E_\ell+s]-F_E[\hat E_\ell] \right) \right], \qquad s=\pm1 , \eeqn which effectively hops electric charge from one site to its nearest neighbor.
The Keldysh term is \beqn S_{w,1{\rm f}} = \gamma_w\int\dd t\sum_\ell 2\cos\tilde A_\ell -\ii\beta K E_\ell\sin\tilde A_\ell , \eeqn  This term depends directly on \(\tilde A_\ell\), and physically it creates a dipole of gauge charges. The strong electric 1-form symmetry is lost with this term. 

We focus on the situation when $\gamma_w t$ is large, and this term dominates at least within certain length scale, meaning again the monopole of $\tilde A_\ell$ may be ignored. Then there is an emergent-dual-weak $\U(1)_c$ symmetry corresponding to the approximate conservation of $\tilde B$. But this weak-$\U(1)_c$ cannot be spontaneously broken, because a strong $\cos\tilde A_\ell$ term condenses the vortices of $\phi$, meaning $\phi$ must be in its disordered (symmetric) phase. Therefore \uline{the normal fluid of model-A of $\U(1)^{(1)}_e$ is the weakly$^\ast$-symmetric phase of $\U(1)_c$}.

If we further assume the weak electric 1-form symmetry has WT-SSB, there is a diffusion of $B$: \beqn \dot B = D_B\nabla^2B + \nabla \times \vect\eta_A. \label{diff2dgaugeWT} \eeqn This is precisely the dual of Eq.~\ref{eq:secII_n_diff_noH} arising from $\U(1)_c$-SW-SSB, also the same EOM as Eq.~\ref{eq:B_quadratic_noH}. Therefore \uline{in $2d$ there is a duality between 1-form model-A and 0-form model-F}. 

\end{document}